\documentclass[aps,prl,twocolumn,superscriptaddress,groupedaddress]{revtex4}  
\usepackage{amssymb}   
\usepackage{amsmath}
\usepackage{graphicx}
\usepackage{verbatim}

\def\beq{\begin{equation}}
\def\eeq{\end{equation}}
\def\bsp#1\esp{\begin{split}#1\end{split}}
\def\bea{\begin{eqnarray}}
\def\eea{\end{eqnarray}}

\begin{document}

\begin{flushleft} 
\mbox{SLAC-PUB-17342, MIT-CTP/5072}
\end{flushleft}

\title{Precision Predictions at N$^3$LO for the Higgs Boson Rapidity Distribution at the LHC}

\author{Falko Dulat}
\affiliation{SLAC National Accelerator Laboratory, Stanford University, Stanford, CA 94039, USA}

\author{Bernhard Mistlberger}
\affiliation{Center for Theoretical Physics, Massachusetts Institute of Technology, Cambridge, MA 02139, USA} 

\author{Andrea Pelloni}
\affiliation{Institute for Theoretical Physics, ETH Z\"urich,
  8093 Z\"urich, Switzerland}

\begin{abstract}
We present precise predictions for the Higgs boson rapidity distribution at the LHC in the gluon fusion production mode.
Our approach relies on the fully analytic computation of six terms in a systematic expansion of the partonic differential cross section around the production threshold of the Higgs boson at next-to-next-to-next-to leading order (N$^3$LO) in QCD perturbation theory. 
We observe a mild correction compared to the previous perturbative order and a significant reduction of the dependence of the cross section on the perturbative scale throughout the entire rapidity range.
\end{abstract}

\maketitle

The precision study of the Higgs boson is key to the current and future physics program at the LHC. 
This is reflected by the remarkable accomplishments~\cite{Aaboud:2018ezd,Sirunyan:2018koj} of  the ATLAS and CMS experiment that continue to test the interactions of the Higgs boson after its discovery~\cite{20121,201230}.
The resulting precise determination of the couplings of the Higgs boson to standard model (SM) particles promises to be a crucial test of physics beyond the SM, especially in absence of direct observations of new particles at the LHC.
Our capability to observe small deviations from SM couplings provides key information to test many models that address problems in high energy physics, such as the origin and nature of dark matter.
Particularly, in the advent of the high luminosity phase of the LHC  it is paramount that the experimental precision is met or surpassed by theoretical predictions in order to reap the full benefit of LHC measurements.

The dominant mechanism for the production of a Higgs boson at the LHC is described by the fusion of two gluons, resolved from the incoming protons, into a virtual top quark loop that then radiates the Higgs boson.
Naturally, there is a significant effort by the particle physics community to determine this particular production mode with utmost precision.
It has long been known that the gluon fusion cross section is afflicted by particularly large perturbative Quantum Chromodynamics (pQCD) corrections~\cite{Dawson:1990zj,Graudenz:1992pv,Spira:1995rr,Djouadi:1991tka,Anastasiou2002,Harlander:2002wh,Ravindran:2003um}. 
This has motivated a long running program to compute higher order QCD corrections to the inclusive gluon fusion cross section that culminated in the recent determination of the next-to-next-to-next-to leading order (N$^3$LO) corrections in pQCD~\cite{Anastasiou:2013srw,Anastasiou:2013mca,Anastasiou:2014vaa,Anastasiou:2014lda,Dulat:2014mda,Anastasiou:2015yha,Anastasiou:2015ema,Anastasiou:2016cez,Dulat:2018rbf,Mistlberger:2018etf}.
Taking into account  effects due to the neglected quark masses as well as electro-weak corrections  and appraising residual uncertainties from missing higher order effects, the current state-of-the-art prediction for the Higgs production cross section in gluon fusion was obtained in~\cite{Anastasiou:2016cez,Dulat:2018rbf}.

Due to the astounding experimental progress we are able to go beyond the determination of total rates and ask more detailed questions about the nature of the Higgs boson. 
In particular, it is possible to perform measurements differential in kinematic variables such as the transverse momentum or rapidity of the Higgs boson. 
Currently, precise predictions through next-to-next-to leading order (NNLO) in QCD are available not only for differential Higgs boson observables but also for observables where a Higgs boson is produced in association with a jet~\cite{Boughezal:2015dra,Chen:2016zka,Chen:2018pzu}. 

\vspace{0.5em}
In this letter we report the calculation of the Higgs boson rapidity distribution at N$^3$LO in QCD perturbation theory. 
The rapidity distribution is the only observable that receives genuine corrections at N$^3$LO that are beyond the formal accuracy of cross sections at NNLO for the production of a Higgs boson in association with a jet.
Our calculation of this observable relies on an approximation of N$^3$LO matrix elements by means of an expansion around the production threshold of the Higgs boson. This drastically simplifies the calculation of the amplitudes contributing to the N$^3$LO cross section and was already successfully used in the calculation of the inclusive corrections to Higgs boson production at N$^3$LO~\cite{Anastasiou:2013srw,Anastasiou:2013mca,Anastasiou:2014vaa,Anastasiou:2014lda,Dulat:2014mda,Anastasiou:2015yha,Anastasiou:2015ema,Anastasiou:2016cez}.
Additionally, we work in an effective theory (EFT) where the top quark is considered to be infinitely heavy and its degrees of freedom are integrated out. Recently, in ref.~\cite{Cieri:2018oms}, the rapidity distribution for Higgs production was also approximated at N$^3$LO in the formalism of $q_T$-subtraction~\cite{Catani:2007vq}, 
exploiting the assumption that one of the ingredients (the third order collinear function) is uniform over the entire rapidity range.

\section{Set-Up}
In collinear factorisation, the probability to produce a Higgs boson with a given rapidity $Y$ can be expressed as 

\bea
\label{eq:hadronicxs}
&&\frac{d\sigma_{P\,P\rightarrow H+X}}{d Y} =\hat \sigma_0 \sum_{i,j}\int_0^1 dx_1dx_2 dy_1 dy_2  f_i(y_1)f_j(y_2) \nonumber\\
&&\hspace{0. cm} \times \delta(\tau-x_1x_2y_1y_2)  \delta\left(Y-\frac{1}{2}\log\left(\frac{x_1 y_1}{x_2 y_2}\right)\right) \eta_{ij}(x_1 ,x_2).\hspace{0.5cm} 
\eea

Here, $f_i(y)$ are parton distribution functions (PDFs) and $\eta_{ij}(x_1,x_2)$ are the partonic coefficient functions (PCFs).
The sum runs over all possible combinations of initial state partons and we integrate over the energy fraction of the incoming partons $y_{1/2}$. 
Furthermore, we define $\tau=m_h^2/S$ and $S=(P_1+P_2)^2$ where the $P_i$ are the momenta of the incoming protons and $m_h$ is the Higgs boson mass. We factor out the leading order partonic cross section $\hat \sigma_0$.

The main result of our calculation is the analytic determination of the PCFs in pQCD through N$^3$LO.
\beq
\eta_{ij}(x_1,x_2)=\sum\limits_{i=0}^3\left(\frac{\alpha_S}{\pi}\right)^i \eta_{ij}^{(i)}(x_1,x_2)+\mathcal{O}(\alpha_S^4).
\eeq
We employ the heavy top quark effective theory which allows us to work only with massless partons and couple the Higgs bosons directly to gluons via an effective interaction. 
The required Wilson coefficient, matching the EFT to full QCD, was computed in refs.~\cite{Chetyrkin:1997un,Schroder:2005hy,Chetyrkin:2005ia,Kramer:1996iq,Gerlach:2018hen}.
The PCFs are comprised of squared partonic matrix elements with up to three unresolved partons in the final state integrated over the available phase space.  
The matrix elements through NNLO are known~\cite{GehrmannDeRidder:2012ja}, but we recompute them using our methodology.
The purely virtual matrix elements and matrix elements with one additional parton in the final state were computed in refs.~\cite{Baikov:2009bg,Gehrmann2010,Duhr:2014nda,Gehrmann:2011aa} and we re-derive them for the purpose of this article. 
Our computation is performed in the framework of dimensional regularisation  in the $\overline{\text{MS}}$ scheme and we rely on previously computed splitting functions~\cite{Moch2004,Vogt2004} and $\beta$-function coefficients~\cite{VanRitbergen:1997va,Czakon:2004bu,Baikov:2016tgj,Herzog:2017ohr} to absorb initial state infrared singularities by a standard mass factorisation redefinition of our PDFs and to perform ultra-violet renormalisation.
Our main result relies on a suitable approximation of squared matrix elements with two and three partons in the final state which we discuss below.

\section{Threshold Expansion}
The probability to produce a Higgs boson via gluon fusion at the LHC is strongly correlated with the probability to find a pair of gluons in the colliding protons. 
This gluon luminosity is steeply falling with the center-of-mass energy $s$ of the gluon pair.
This results in an enhancement of the hadronic cross section, when the Higgs boson is produced close to threshold, i.e.~when $s$ equals the mass of the Higgs boson.
This kinematic enhancement was exploited successfully in the past to perform precise approximations of the inclusive Higgs boson production cross section in terms of a systematic expansions around the production threshold~\cite{Harlander:2002wh,Anastasiou:2015ema}. 
In this limit the threshold parameter $z=m_h^2/s=x_1 x_2$ tends to one and an expansion can be performed around $\bar z=1-z=0$.

In ref.~\cite{Dulat:2017prg} we demonstrated that the rapidity distribution at NNLO can be approximated to a high degree of precision using a threshold expansion.
Furthermore, we already obtained the first two terms in the threshold expansion of the PCFs at N$^3$LO. 
In this article we go beyond this result and obtain in total the first six terms of the expansion in $\bar z$. 
We achieve this by following the strategies outlined in ref.~\cite{Dulat:2017prg} based on integrand expansions of Higgs differential cross sections~\cite{Dulat:2017aa,Anastasiou:2013mca,Anastasiou:2015yha,Anastasiou:2013srw}, which generalise the techniques employed at NNLO~\cite{Anastasiou:2002qz,Anastasiou:2003yy,Anastasiou2004a}. 
The result is a PCF differential in the transverse momentum and rapidity of the Higgs boson. 
In order to obtain our $\eta_{ij}^{(3)}(x_1,x_2)$ we analytically integrate out the extra degree of freedom corresponding to the transverse momentum.

In the variables $x_1$ and $x_2$ the threshold expansion can be performed by introducing a formal expansion parameter $\delta$ such that
\beq
\label{eq:thresh}
\bar x_1\rightarrow \delta \bar x_1\frac{1-\bar x_2}{1-\delta \bar x_2},\hspace{0.5cm}\bar x_2\rightarrow \delta \bar x_2.
\eeq
Here, $\bar x_i=1-x_i$. 
The expansion parameter $\delta$ is chosen exactly such that  each term in the expansion around $\delta =0$ of our PCF corresponds exactly to one term in the expansion in $\bar z$.

\section{Exploiting the Divergence Structure}
The bare PCFs at N$^3$LO, arising from the calculation of contributing squared matrix elements, in $d=4-2\epsilon$ dimensions take the form 
\bea
\label{eq:bareeta}
\eta_{ij,\,\text{bare}}^{(3)}(\bar x_1,\bar x_2)&=&\eta_{ij,\,\text{virt.}}^{(3)}\delta(\bar x_1)\delta(\bar x_2)\\
 &+&\sum_{n,m =1}^3  \bar x_{1}^{-1-m \epsilon} \bar x_{2}^{-1-n\epsilon} \eta_{ij,\,\text{bare}}^{(3,m,n)}(\bar x_1,\bar x_2).\nonumber
\eea
The term $\eta_{ij,\,\text{virt.}}^{(3)}\delta(\bar x_1)\delta(\bar x_2)$ corresponds to the purely virtual contributions with a leading divergence of $1/\epsilon^6$. 
The functions $ \eta_{ij,\,\text{bare}}^{(3,n,m)}(\bar x_1,\bar x_2)$ are holomorphic around $\bar x_i=0$ and contain fourth order poles in $\epsilon$.
In order to expand the PCFs in the dimensional regulator we perform a standard expansion of singular factors in terms of delta functions $\delta(\bar x_i)$ and plus-distributions 
$\left[L^n_i/\bar{x}_i\right]_+$
with $L_i=\log(\bar x_i)$.

We obtain our finite, renormalised N$^3$LO coefficient function by combining the bare PCF with a suitable mass factorisation and ultraviolet renormalisation counter term $CT^{(3)}_{ij}$. 
This counter term equally contains distributions and renders the renormalised PCFs finite,
\bea
\label{eq:polecancellation}
\eta_{ij}^{(3)}( x_1, x_2)&=&\lim\limits_{\epsilon\rightarrow 0}\left[\eta_{ij,\,\text{bare}}^{(3)}( x_1, x_2)+CT^{(3)}_{ij}( x_1, x_2)\right]\nonumber\\
&=&\sum\limits_{k,l} D_k(\bar x_1) D_l(\bar x_2) \eta_{ij,(k,l)}^{(3)}( x_1, x_2).
\eea
In the second line in the above equation we isolate the structures that are singular in the limit $\bar{x}_i\to0$ into the functions $D_k(\bar x_i)$ such that the coefficients $\eta_{ij,(k,l)}^{(3)}( x_1, x_2)$ are either real numbers or holomorphic functions in the limit.
The non-holomorphic function $D_k(\bar x_i)$ corresponds to the $k^{th}$ entry of the following list of 12 possible structures that can appear in the cross section through N$^3$LO,
\beq
 \left\{\delta(\bar x_i),\left[\frac{L^0_i }{\bar x_i}\right]_+,\dots,\left[\frac{L^5_i}{\bar x_i}\right]_+,L^0_i ,\dots L^4_i\right\}.
\eeq

The fact that all explicit poles in the dimensional regulator have to cancel among the different contributions in eq.~\eqref{eq:polecancellation}, allows us to derive relations among the various bare partonic coefficients in eq.~\eqref{eq:bareeta} and the counter term.
Using the fact that only the known, genuine two loop contributions can produce bare coefficient functions contributing to the $n=1$ or $m=1$ terms in eq.~\eqref{eq:bareeta}, this becomes a powerful tool to determine many of the coefficients $\eta_{ij,(k,l)}^{(3)}( x_1, x_2)$ exactly.

All coefficients of terms proportional to two distributions were already computed in ref.~\cite{Dulat:2017prg} and can also be deduced from the inclusive cross section at threshold (c.f. ref.~\cite{Anastasiou:2014vaa,Li:2014afw}) as was done in ref.~\cite{Ahmed:2014uya}. 
Furthermore, we observe that if we consider only the leading power term in either one of the $\bar x_i$, a reduced number of exponents contributes to eq.~\eqref{eq:bareeta}, such that $n\leq m$, which further constrains our system of equations. 
Using these relations, we were able to determine all coefficients in eq.~\eqref{eq:polecancellation} exactly in $\bar{x}_i$, except for the terms, 
\bea
\label{eq:missing}
&&\eta_{ij,\,\text{missing}}^{(3)}( x_1, x_2)=\Big[ \delta(\bar x_1) \log(\bar x_2) \eta_{ij,(1,9)}^{(3)}(0, x_2)   \\
&&\hspace{0.5cm}+\delta(\bar x_1) \eta_{ij,(1,8)}^{(3)}(0, x_2) +\left[\frac{1}{\bar x_1}\right]_+ \eta_{ij,(2,8)}^{(3)}(0, x_2)\nonumber\\
&&\hspace{0.5cm}+\log(\bar x_2) \eta_{ij,(8,9)}^{(3)}( x_1, x_2)\Big]+\Big[( x_1 \leftrightarrow  x_2)\Big] \nonumber\\
&&\hspace{0.5cm}+\eta_{ij,(8,8)}^{(3)}( x_1, x_2)+\log(\bar x_1) \log(\bar x_2) \eta_{ij,(9,9)}^{(3)}( x_1, x_2) .\nonumber
\eea
While the above terms could not be determined exactly from our current knowledge of unexpanded matrix elements, we obtained them via a threshold expansion as described above.
Notice, that the above contributions contain maximally one power of a logarithm that is enhanced as $\bar x_i \rightarrow 0$. 

In our final approximation of the PCF we choose to re-organise the terms without any distributions such that all terms proportional to threshold logarithms $\log^i(\bar z)$ with $i\geq 3$ are maintained exactly. 
We approximate terms with lower powers of threshold logarithms using the threshold expansion as discussed above.
Note, that the relations among the different components of the PCF provide a highly non-trivial consistency check on the results from our threshold expansion.

The partonic coefficient functions also depend explicitly on logarithms of the perturbative scale $\mu$ and we can rearrange them as,
\beq
\eta_{ij}^{(3)}( x_1, x_2)=\sum_{l=0}^3  \eta_{ij}^{(3,l)}( x_1, x_2) \log^l\left(\frac{m_h^2}{\mu^2}\right).
\eeq
Naturally, the functions $ \eta_{ij}^{(3,l)}( x_1, x_2)$ can be decomposed into distribution-valued or logarithmically enhanced terms as above.
However, the coefficients with $l\geq1$ can be derived exactly from lower order cross sections by solving the DGLAP evolution equations.
Consequently, we determine them exactly, with one exception:
The derivation of the non-distribution valued, non-logarithmically enhanced term of the coefficient of $\log\left(m_h^2/\mu^2\right)$ involves rather cumbersome convolution integrals. 
We approximate this particular term using a threshold expansion which modifies our approximation at terms beyond the claimed formal accuracy.

\section{Matching to the Inclusive Cross Section}
The inclusive PCF for Higgs boson gluon fusion production at N$^3$LO was computed exactly in ref.~\cite{Mistlberger:2018etf}.
It is a one parameter function of the threshold variable $z$.
By performing the variable transformation \mbox{$\{\bar x_1=\frac{(1-\bar x)\bar z}{1-\bar x\bar z},\bar x_2=\bar x\bar z\}$}  we can relate our differential PCF to the inclusive one,
\beq
\label{eq:incint}
\eta_{ij}^{(3),\text{inc.}}(z)=\int_0^1\frac{\bar z d\bar{x}}{(1-\bar z \bar x)} \eta_{ij}^{(3)} \left(\frac{(1-\bar x)\bar z}{1-\bar x\bar z},\bar x\bar z\right)\,.
\eeq
The above relation provides an enormously stringent check on our partonic coefficient functions. 
Indeed, our threshold expansion agrees with the threshold expansion of the inclusive partonic coefficient function for all computed orders.

Furthermore, eq.~\eqref{eq:incint} allows us to modify our differential partonic coefficient functions by terms of higher order in the threshold expansion such that the exact inclusive cross section is automatically obtained if the integral over the rapidity distribution is performed,
\bea
\label{eq:ourapprox}
&& \eta_{ij}^{(3) ,\text{matched}}( x_1,  x_2)=\eta_{ij}^{(3) ,\text{app.}}( x_1,  x_2)\\
&&\hspace{0.5cm}+\frac{x_1+x_2}{2(1-x_1 x_2)}\left[\eta_{ij}^{(3),\text{inc}}(x_1 x_2)- \eta_{ij}^{(3),\text{inc, app.}}(x_1 x_2)\right].\nonumber
\eea
Here, $\eta_{ij}^{(3) ,\text{app.}}$ corresponds to the approximation of the PCF obtained as described in the previous sections and $\eta_{ij}^{(3),\text{inc, app.}}$ is its inclusive counterpart obtained by virtue of eq.~\eqref{eq:incint}.
Furthermore, $\eta_{ij}^{(3),\text{inc}}$ is the inclusive partonic coefficient function obtained in ref.~\cite{Mistlberger:2018etf}. 
The term in the square bracket of eq.~\eqref{eq:ourapprox} contains therefore only terms that are higher order in the threshold expansion than those obtained as described above.
This modification of the PCF ensures that if the inclusive integral over the Higgs boson rapidity is performed the correct cross section is obtained for each partonic center of mass energy. The approximation derived in eq.~\eqref{eq:ourapprox} will be the basis for our numerical results presented below.

\section{Phenomenological Results}
In the previous sections we derive an analytic approximation to the PCF for the Higgs boson cross section differential in the rapidity through N$^3$LO in QCD.
We now use MMHT2014 PDFs~\cite{Thorne:2015rch} to derive predictions for hadronic Higgs boson rapidity distribution at the LHC with a centre of mass energy of $13$ TeV by means of eq.~\eqref{eq:hadronicxs}. We implement our coefficient functions into a private C++ code and use LHAPDF~\cite{Buckley:2014ana} to perform the $\mu^2$ evolution of the PDF grids and evaluate them with a private grid interpolator. The Cuba library~\cite{Hahn:2004fe} is used to perform the numerical integration over the momentum fractions of the partons.

\begin{figure*}[!t]
\begin{center}
\includegraphics[width=0.48\textwidth]{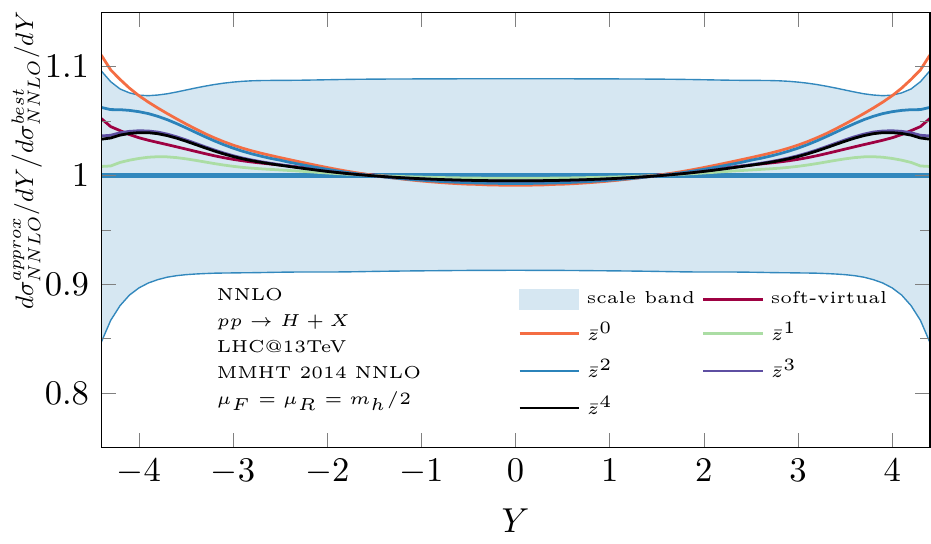}
\includegraphics[width=0.48\textwidth]{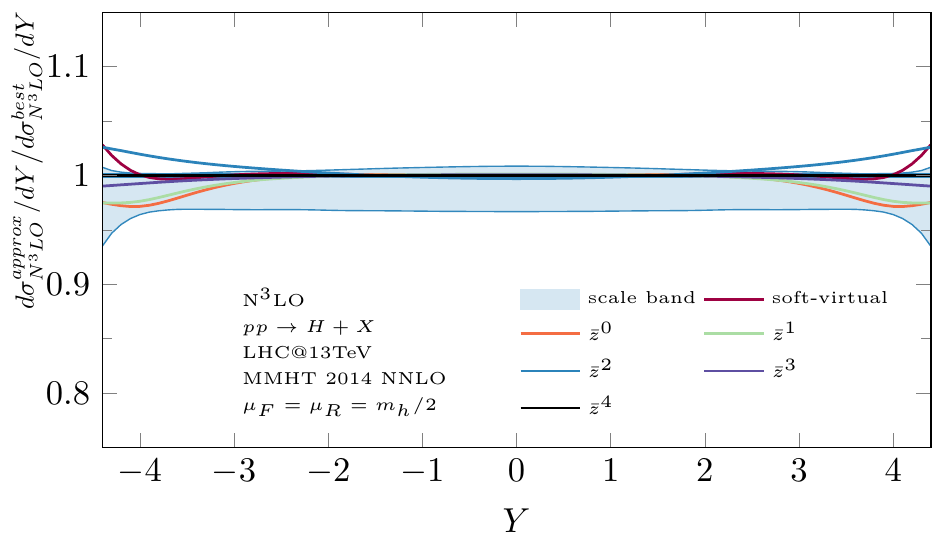}
\end{center}
\caption{
\label{fig:approx}
Approximate Higgs boson rapidity distribution with threshold expansion truncated at different orders. 
The left panel shows the ratio of the approximate NNLO to the exact result, the right panel shows the approximate N$^3$LO result to the best prediction obtained in this work. }
\end{figure*} 
As validation, we first derive the NNLO analogue of the approximation of the PCF used at N$^3$LO and show the resulting predictions in the left panel of fig.~\ref{fig:approx} normalised to the exact rapidity distribution through NNLO with a central scale of $\mu=m_h/2$. The blue band corresponds to the cross section obtained by varying the common scale $\mu$ in the interval $[m_h/4,m_h]$.
The coloured lines show the cross section obtained by truncating the threshold expansion in our approximation at different orders.
We observe that our approximation describes the NNLO rapidity distribution very well for central rapidities $(|Y|<3)$ and even performs fine for larger rapidities. Deterioration of the threshold approximation at larger rapidities can be expected as on average the final state of the scattering process is more energetic,~i.e. further from the production threshold.
Including an increasing number of terms systematically improves the approximation.
We also observe that all rapidity distributions obtained from truncated threshold expansions fall well within the scale variation band of the exact NNLO cross section.

In the right panel of fig.~\ref{fig:approx} we show predictions for the N$^3$LO rapidity distribution truncating the threshold expansion at different orders normalised to our best approximation. Similarly to the case at NNLO, including more terms in the expansion systematically stabilises our approximation.
Central rapidities are remarkably stable under the inclusion of more and more expansion terms. In particular, all truncated approximations are once again contained within the scale variation band for central rapidities. 
We explored relaxing some of the ingredients of our approximation (less exact distributions or no matching to the exact inclusive cross section) which amounts to a modification of terms beyond those computed in our threshold expansion and find only slight variation in our prediction. 
For example, basing our calculation purely on a threshold expansion with six terms, underestimates the inclusive cross section by $0.25\%$ and only slightly varies the shape of the rapidity distribution.
Similarly, we checked that a simple reweighting of the threshold-expanded N$^3$LO rapidity distribution to the exact inclusive cross section at N$^3$LO produces results that are very close to our best prediction including the matching procedure according to eq.~\eqref{eq:incint}.
We observe that at NNLO we approximate the exact PCF to better than one percent for $|Y|<2$ and better than two percent for $|Y|<3$. 
In order to be conservative we estimate that our prediction is at the same level of precision relative to the exact result at N$^3$LO.

\begin{figure}[!t]
\begin{center}
\includegraphics[width=0.48\textwidth]{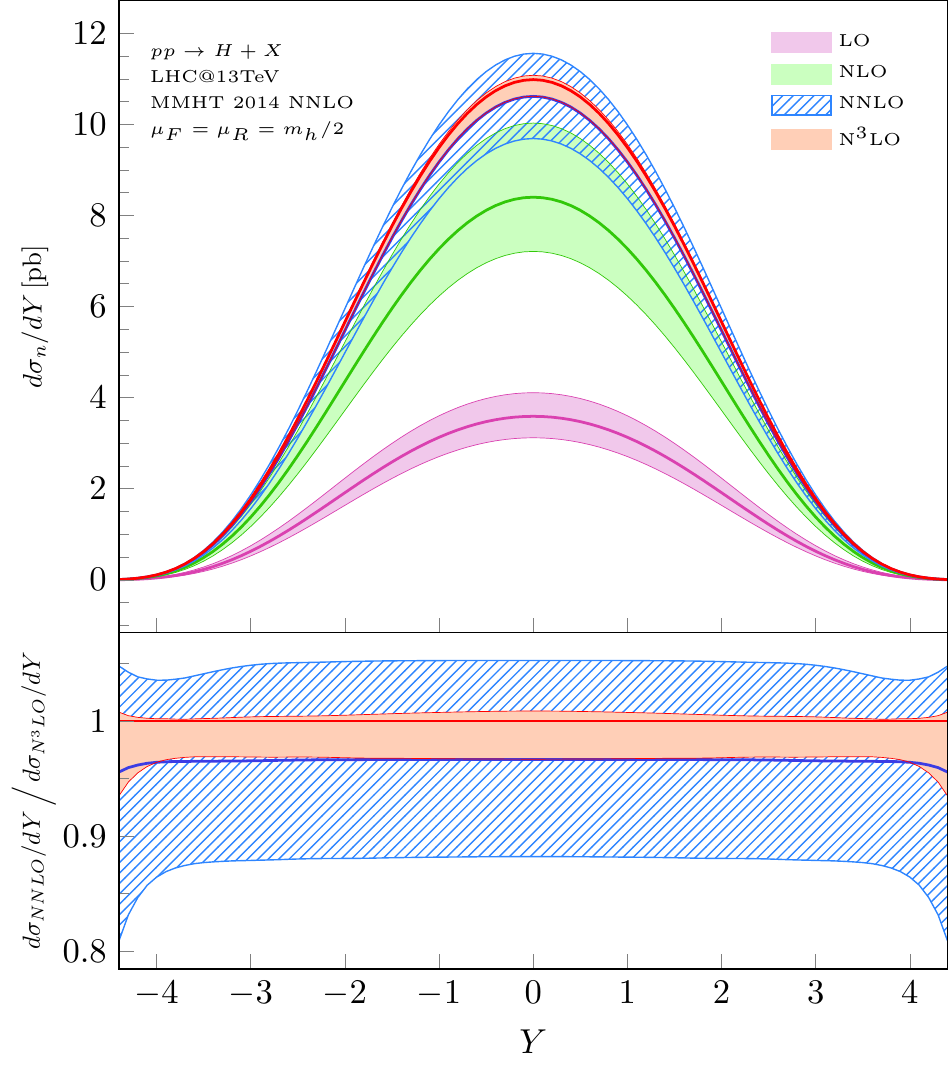}
\end{center}
\caption{
\label{fig:rap}
The Higgs boson rapidity distribution at different orders in perturbation theory. 
The lower panel shows the N$^3$LO and NNLO predictions normalised to the N$^3$LO prediction for $\mu=m_h/2$.
}
\end{figure} 

In fig.~\ref{fig:rap} we show the rapidity distribution of the Higgs boson truncated at different orders in QCD perturbation theory. 
Our newly derived N$^3$LO predictions display a stabilisation of the perturbative series as well as a drastic reduction of the size of perturbative scale dependence. 
We observe that the ratio of the rapidity distribution at N$^3$LO relative to NNLO is uniform over the entire range of Higgs boson rapidities. Consequently, the N$^3$LO rapidity distribution can be reproduced to very high accuracy by rescaling the NNLO prediction by the inclusive N$^3$LO k-factor. 
Our findings for the central value and scale variation of the rapidity distribution are in agreement with the result presented in ref.~\cite{Cieri:2018oms}. 
At very large rapidities the authors of ref.~\cite{Cieri:2018oms} observe a slight deviation from an entirely uniform N$^3$LO correction but our predictions are still compatible within uncertainties.

To conclude, in this article we have obtained theoretical predictions for the Higgs boson rapidity distribution at significantly improved levels of precision. 
The scale variation of the N$^3$LO cross section 
for $|Y|<3$ is reduced to $[-3.4\%,+0.9\%]$ and we estimate the uncertainty due to missing higher orders in the threshold expansion to be less than $1\%$ for $|Y|<2$ and less than $2\%$ for $|Y|<3$.
Our result has direct implications for the LHC phenomenology program and represents a mile stone in the field of perturbative QCD. 
We expect the result of this work to be the corner stone of future fully differential Higgs boson phenomenology.

{\bf Acknowledgements:}
We are grateful to Alexander Huss for useful discussions. We thank Babis Anastasiou, Thomas Gehrmann, Stefan Hoeche and Giulia Zanderighi for useful comments on the manuscript.
The research of FD is supported by the U.S. Department of Energy
(DOE) under contract DE-AC02-76SF00515.
AP is supported by the European Commission through the ERC grant pertQCD.
BM is supported by the Pappalardo fellowship and  was supported by the European Comission through the ERC Consolidator Grant HICCUP (No. 614577).


\bibliography{Bib}

\end{document}